\documentclass[aps,prl,twocolumn,groupedaddress,showpacs]{revtex4}

\usepackage{graphicx}

\begin{document}

\title{Propagation of Bose-Einstein condensates in a magnetic waveguide}

\author{A.E.~Leanhardt, A.P.~Chikkatur, D.~Kielpinski, Y.~Shin, T.L.~Gustavson, W.~Ketterle, D.E.~Pritchard}

\homepage[URL: ]{http://cua.mit.edu/ketterle_group/}

\affiliation{Department of Physics, MIT-Harvard Center for
Ultracold Atoms, and Research Laboratory of Electronics,
Massachusetts Institute of Technology, Cambridge, Massachusetts,
02139}

\date{\today}

\begin{abstract}
Gaseous Bose-Einstein condensates of $2-3 \times 10^6$ atoms were
loaded into a microfabricated magnetic trap using optical
tweezers. Subsequently, the condensates were released into a
magnetic waveguide and propagated 12~mm. Single-mode propagation
was observed along homogeneous segments of the waveguide.
Inhomogeneities in the guiding potential arose from geometric
deformations of the microfabricated wires and caused strong
transverse excitations. Such deformations may restrict the
waveguide physics that can be explored with propagating
condensates.
\end{abstract}

\pacs{03.75.Fi, 03.75.Be, 39.20.+q}

\maketitle

Progress in the field of atom optics depends on developing
improved sources of matter waves and advances in their coherent
manipulation.  Extended samples of coherent atoms were introduced
to the field in 1995 with the realization of Bose-Einstein
condensation in dilute alkali vapors~\cite{ISW99}. Early success
in coherently controlling condensates was achieved by applying
optical and radio frequency fields~\cite{ISW99}. Further efforts
to increase the repertoire of coherent manipulation techniques for
ultracold atomic vapors pursued miniaturizing the current carrying
structures often used in their study~\cite{WEL95,TOZ99}. Following
the successful trapping and guiding of thermal atoms using
self-supported miniature wires~\cite{FGZ98,DCS99,KHR00} and
substrate-supported microfabricated wire
arrays~\cite{RHH99,MAG99,DLL00,FKC00}, recent experiments merged
wiretraps on the millimeter scale~\cite{GCL02} and microfabricated
electronic devices~\cite{OFS01,HHH01} with Bose-Einstein
condensation.  This has opened up a front on which further
techniques for coherent condensate transport and manipulation can
be explored.  While previous demonstrations of condensate guiding
with optical potentials will be limited fundamentally by
diffraction~\cite{BBD01}, fundamental limitations to guiding
condensates with microfabricated surfaces are not expected for an
atom-surface separation in excess of 1~$\mu$m~\cite{HEW99}.

In this Letter, we demonstrate that a Bose-Einstein condensate
(BEC) transported with optical tweezers can be transferred into a
microtrap on a substrate.  Such condensates contained five times
more atoms than those created in a similar microtrap~\cite{OFS01}.
We released the BEC from the microfabricated magnetic trap into a
single-wire magnetic waveguide and studied its propagation.
Condensates were observed to propagate 12~mm before exiting the
field-of-view of our imaging system. The atoms propagated with a
longitudinal kinetic energy in excess of the transverse
confinement energy of the waveguide. This regime is important for
atom interferometry because it reduces sensitivity to
imperfections in the interferometer and the probability of atoms
being reflected from interferometric elements.

We observed single-mode (excitation-less) BEC propagation along
homogeneous segments of the waveguide.  Transverse excitations
were created in condensates propagating through perturbations in
the guiding potential. These perturbations resulted from geometric
deformations of the current carrying wires on the substrate. Finer
imperfections were observed when trapped condensates were brought
closer to the surface of the microchip as evidenced by the
longitudinal fragmentation of the cloud.

The apparatus used for this experiment was recently described in
Ref.~\cite{GCL02}.  Condensates containing over $10^7$ $^{23}$Na
atoms were created in the $|F,m_F\rangle=|1,-1\rangle$ state in a
macroscopic Ioffe-Pritchard magnetic trap using forced radio
frequency (RF) evaporative cooling~\cite{KDS99}. An optical
tweezers beam then captured the BEC and transported it to within
500~$\mu$m of the microfabricated structures.  The microchip was
contained in an auxiliary ``science'' chamber isolated from the
BEC production chamber by bellows and a gate valve. The
translation distance of the tweezers was $\approx 30$~cm and was
traversed in 1250~ms. This quick transfer allowed for reliable
delivery of condensates containing $2-3 \times 10^6$ atoms into
the science chamber.

\begin{figure}
\begin{center}
\includegraphics{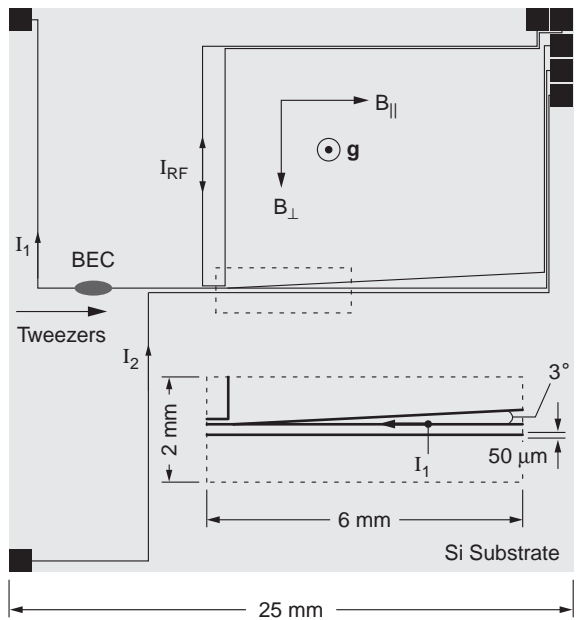}
\caption{Microfabricated magnetic trap and waveguide. Optical
tweezers loaded a Bose-Einstein condensate into the Z-wire trap
formed by currents $I_1$ and $I_2$ in conjunction with the
magnetic bias field $B_\bot$. Lowering $I_2$ to zero released the
condensate into a single-wire magnetic waveguide. Atom flow was
from left to right. The inset shows the widening of the waveguide
wire in the region where another wire merges with it at a small
angle. The only current flowing in the inset is $I_1$. The
condensate was trapped above the plane of the paper and the
gravitational force, \textbf{g}, points out of the page. All
microfabricated features are drawn to scale.\label{f:chip}}
\end{center}
\end{figure}

A schematic of the microchip onto which the BEC was loaded is
shown in Fig.~\ref{f:chip}.  The microfabricated wires lie on a
600~$\mu$m thick silicon substrate.  They are 50~$\mu$m wide and
electroplated with copper to a height of 10~$\mu$m.  The minimum
separation distance between wires is 50~$\mu$m (100~$\mu$m
center-to-center). The silicon wafer was mounted on an aluminium
block for heat conduction purposes. Macroscopic leads that
extended to a vacuum feedthrough were attached to the wafer using
a gap welding technique.

The BEC was initially loaded into a Z-wire trap~\cite{RHH99}
formed by currents $I_1$ and $I_2$ along with an orthogonal
magnetic bias field $B_\bot$.  Typical loading conditions were
$I_1=I_2=1200$~mA and $B_\bot=5.4$~G, corresponding to a
separation of 450~$\mu$m between the BEC and the microchip. A
longitudinal bias field, $B_{||}$, tailored the transverse trap
frequency while nominally keeping the trap center in the same
location relative to the surface of the microchip. Transfer
efficiency from the optical tweezers to the Z-wire trap was near
unity and BEC lifetimes over 10~s were observed with the
application of a RF shield~\cite{KDS99} produced by the current
$I_{RF}$ on an auxiliary wire as shown in Fig.~\ref{f:chip}.

The BEC was transferred into the waveguide by linearly ramping the
current $I_2$ to zero in 250~ms. The atoms were smoothly
accelerated into the waveguide by the remaining endcap of the
Z-wire trap. Downstream, the effect of this endcap was negligible
and we observed BEC propagation at a constant velocity of 3.0~cm/s
after a propagation distance of 4~mm for $I_1=1200$~mA. Upon
releasing the BEC from the Z-wire trap, its longitudinal velocity
was controlled by applying an external magnetic field gradient of
variable amplitude for a fixed time.  The field gradient was
linearly ramped up and down over 6.5~ms to prevent creating
excitations and was held constant for 52~ms. With modest gradients
of $0-0.6$~G/cm, we were able to vary the atomic velocity over the
range $3.0-6.6$~cm/s corresponding to longitudinal kinetic
energies in the range $25-120$~kHz.

Releasing the BEC into the waveguide produced little variation in
its transverse confinement and position relative to the microchip.
The longitudinal potential experienced by the propagating atoms
was determined by the local magnetic field (due to the Zeeman
interaction) and vertical position of the guide center (due to the
gravitational interaction).  For $^{23}$Na atoms in the
$|1,-1\rangle$ state, the Zeeman energy is 700~Hz/mG and the
gravitational energy is 560~Hz/$\mu$m.

Perturbations to the guiding potential arise from geometric
deformations of the current carrying wires on the substrate. The
extent to which such deformations alter the potential experienced
by the atoms depends on the atom-wire separation distance, $r$,
longitudinal extent of the perturbation, $\ell$, wire width, $w$,
and wire height, $h$. Under our guiding conditions ($r \gg w,h$)
the waveguide potential only responds to changes in the centroid
of the current density. In general, only wire deformations with
$\ell \gtrsim r$ will significantly perturb the guiding potential
due to solid angle considerations.

Such a deformation ($\ell \approx 1$~mm, $r=450$~$\mu$m) is
depicted in the inset to Fig.~\ref{f:chip}. As the wire width
varies during the bifurcation process, the centroid of the current
density will shift in the plane of the microchip. This shift will
be mirrored by a shift in the trajectory of the guide in a plane
parallel to that of the microchip~\cite{height}.  Such a shift
causes the guide axis to make an angle, $\theta$, with respect to
its nominal trajectory along which $B_{||}$ and $B_\bot$ are
aligned parallel and perpendicular respectively. As a result, the
effective parallel, $B'_{||}$, and perpendicular, $B'_\bot$,
magnetic fields are found by the rotation
\begin{eqnarray}
    \label{e:bparaeff}
    B'_{||} & = & B_{||} \cos \theta - B_\bot \sin \theta,\\
    \label{e:bperpeff}
    B'_\bot & = & B_\bot \cos \theta + B_{||} \sin \theta,
\end{eqnarray}
where $-\pi/2 \leq \theta \leq \pi/2$.  $\theta$ is taken to be a
positive(negative) angle for the specific case of atoms
entering(exiting) the waveguide perturbation depicted in the inset
to Fig.~\ref{f:chip}.

Changes in the effective parallel and perpendicular magnetic
fields result in variations in the longitudinal potential
experienced by the atoms due to the Zeeman and gravitational
interactions respectively.  Variations in the effective parallel
magnetic field are given by (for small angles)
\begin{equation}
    \label{e:db}
    \Delta B'_{||} = -B_\bot \sin \theta.
\end{equation}
Thus, atoms entering(exiting) the perturbed guiding region will
encounter a magnetic potential well(barrier). Furthermore, changes
in the effective perpendicular bias field cause the atom-substrate
distance to vary by (for small angles)
\begin{equation}
    \label{e:dr}
    \Delta r = -r\frac{\Delta B_\bot}{B_\bot} = - r \frac{B_{||}}{B_\bot} \sin \theta.
\end{equation}
Thus, the guide center entering(exiting) the perturbed guiding
region will shift towards(away from) the surface of the microchip
and atoms will encounter a gravitational potential barrier(well).
For the waveguide parameters used in this work, perturbations to
the guiding potential are dominated by longitudinal variations in
the Zeeman energy.

\begin{figure}
\begin{center}
\includegraphics{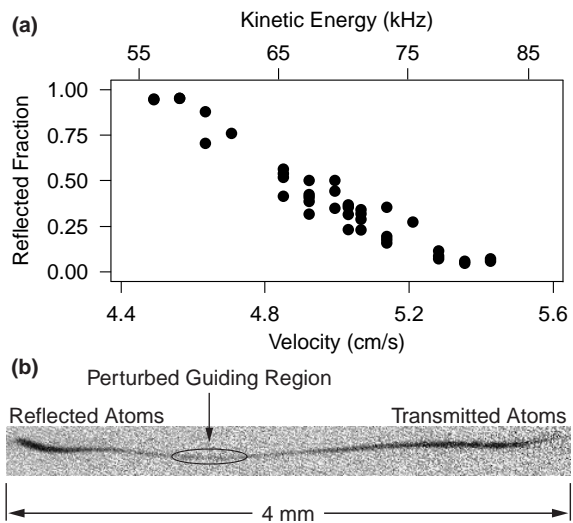}
\caption{Bose-Einstein condensates scattering from the waveguide
perturbation shown in the inset to Fig.~\ref{f:chip}. (a) The
fraction of atoms reflected as a function of incident velocity.
(b) Absorption image after 8~ms of ballistic expansion of a
partially transmitted condensate.  The incident longitudinal
velocity and kinetic energy were 5.0~cm/s and 70~kHz respectively.
Waveguide parameters for all data were $I_1 = 1200$~mA, $B_\bot =
5.4$~G, $B_{||} = 3.6$~G, and $\omega_\bot = 2\pi \times 84.5$~Hz.
\label{f:reflection}}
\end{center}
\end{figure}

The effects of the perturbation depicted in the inset to
Fig.~\ref{f:chip} on BEC propagation were studied by varying the
incident velocity of the atoms sent into the region. The BEC
entered the perturbed guiding region after $\geq 100$~ms of
propagation, well after the external gradient used to accelerate
the atoms was turned off. Condensates travelling below 4.5~cm/s
(55~kHz kinetic energy) were totally reflected from the
perturbation, while clouds at speeds above 5.4~cm/s (80~kHz
kinetic energy) were entirely transmitted. Fig.~\ref{f:reflection}
shows the results for intermediate atomic velocities.

The local magnetic field in the perturbed guiding region was
measured by driving RF spin-flip transitions that removed atoms
from the guide~\cite{KDS99}. It was found that upon
entering(exiting) the perturbed region the magnetic bottom of the
guiding potential decreased(increased) by $50 \pm 10$~kHz. The
signs of these shifts are consistent with those predicted by
Eq.~\ref{e:db}.  The magnitude of the shift in the Zeeman energy
upon exiting the guide is also consistent with the onset of
transmission through the perturbed guiding region as shown in
Fig.~\ref{f:reflection}(a). From Eq.~\ref{e:db} with $B_\bot =
5.4$~G, the maximum angular deviation, $\theta_m$, of the
waveguide trajectory necessary to produce a 50~kHz perturbation to
the Zeeman energy is $\theta_m = 13$~mrad. The corresponding
vertical displacement of the guide center from Eq.~\ref{e:dr} with
$B_\bot = 5.4$~G, $B_{||}=3.6$~G, and $\theta_m = 13$~mrad is
4~$\mu$m. This yields a gravitational potential variation of
2.2~kHz, which is small compared to the 50~kHz Zeeman energy
shifts associated with the perturbation.

\begin{figure}
\begin{center}
\includegraphics{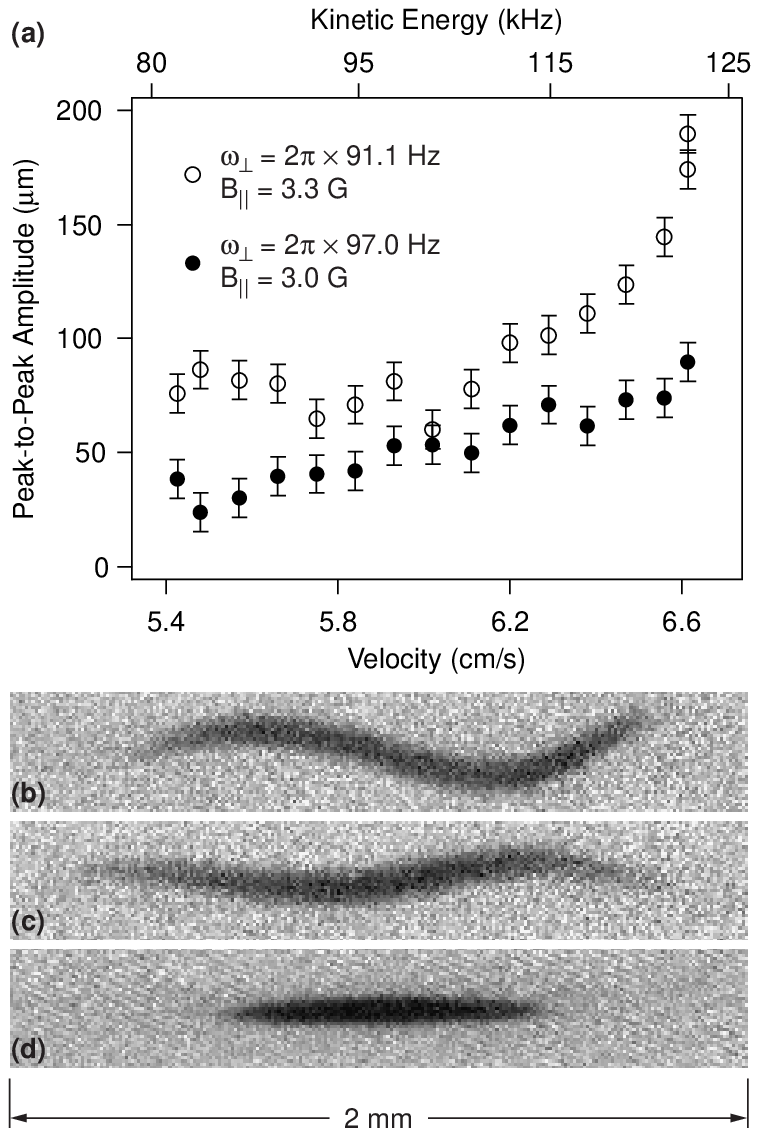}
\caption{Velocity dependence of Bose-Einstein condensate
excitations after propagating through the waveguide perturbation
depicted in the inset to Fig.~\ref{f:chip}. (a) Peak-to-peak
amplitudes of condensate excitations after 15~ms of ballistic
expansion versus velocity, $v$. Absorption images of condensates
with $v=6.5$~cm/s after 15~ms of ballistic expansion for (b)
$\omega_\bot = 2\pi \times 91.1$~Hz and $B_{||} = 3.3$~G and (c)
$\omega_\bot = 2\pi \times 97.0$~Hz and $B_{||} = 3.0$~G. (d)
Typical condensate after 10~ms of ballistic expansion ($N \approx
2 \times 10^6$ atoms, $v=6.3$~cm/s, $\omega_\bot = 2\pi\times
84.5$~Hz, and $B_{||} = 3.6$~G) just prior to entering the
perturbed guiding region. The wire current, $I_1 = 1200$~mA, and
perpendicular bias field, $B_\bot = 5.4$~G, were held fixed for
all data. All transmitted condensates propagated 4~mm beyond the
perturbed guiding region before being imaged.
\label{f:excitations}}
\end{center}
\end{figure}

In the propagation regime where the longitudinal kinetic energy of
the BEC is large compared to the transverse confinement energy of
the guide, perturbations involving transverse shifts in the guide
trajectory are expected to transversely excite incident
condensates. Fig.~\ref{f:excitations} depicts such transverse
excitations for condensates transmitted through the waveguide
perturbation depicted in the inset to Fig.~\ref{f:chip}.  The
incident BEC velocity was in the range $5.4-6.6$~cm/s
corresponding to longitudinal kinetic energies in the range
$80-120$~kHz. Excitations were characterized by the peak-to-peak
amplitude of the transverse displacement of the BEC. The imaging
axis only provided sensitivity to transverse excitations in the
plane normal to the surface of the microchip.

Fig.~\ref{f:excitations}(a) shows a clear increase in the
excitation amplitude with increasing velocity.  Furthermore, the
excitation amplitude for fixed incident atomic velocity should
increase when either the ratio of longitudinal kinetic energy to
transverse confinement energy or the size of the perturbation
increases. By increasing $B_{||}$ we simultaneously lowered the
trap frequency and, according to Eq.~\ref{e:dr}, increased the
vertical displacement of the guide center. This resulted in an
observed increase in excitation amplitude for all velocities
studied.

The nearly sinusoidal nature of the BEC excitations shown in
Fig.~\ref{f:excitations}(b) and (c) indicates that we were
primarily exciting the dipole mode of the BEC. Further evidence
for this comes from the fact that we observed little variation in
the excitation amplitude as a function of the propagation time
after exiting the perturbed guiding region. Conversely,
transmitted condensates showed strong signs of higher order
excitations when $B_{||}$ was increased beyond 3.6~G corresponding
to $\omega_\bot \leq 2\pi\times 84.5$~Hz. The absorption images
deviated visibly from a smooth sinusoidal shape.  In addition, the
measured excitation amplitude depended strongly on propagation
time indicting the phasing and dephasing of several excitation
modes.

\begin{figure}
\begin{center}
\includegraphics{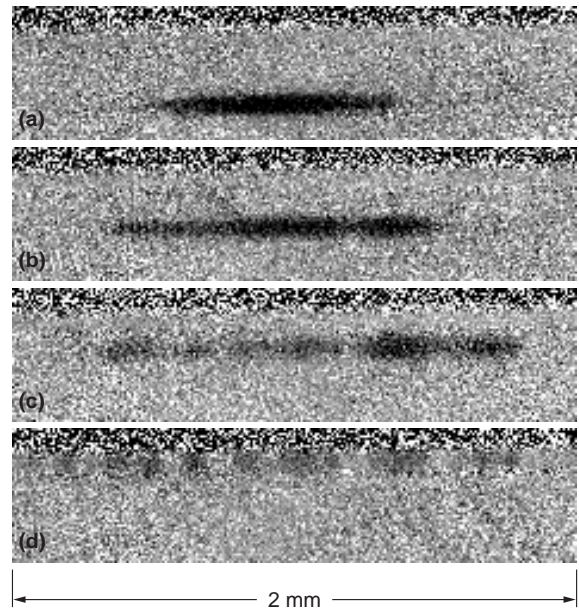}
\caption{Fragmented structure of Bose-Einstein condensates brought
close to the substrate.  Absorption images after 5~ms of ballistic
expansion from the Z-wire trap shown in Fig.~\ref{f:chip}.  The
atom-substrate separation was (a) 190~$\mu$m, (b) 145~$\mu$m, (c)
100~$\mu$m, and (d) 55~$\mu$m.  For all images, the condensate
started in a trap with $I_1=I_2=540$~mA, $B_\bot = 5.4$~G, and
$B_{||}=0.3$~G corresponding to an atom-substrate separation of
200~$\mu$m.  The condensate was translated towards the substrate
by lowering the wire currents linearly over 500~ms. The atoms were
held at their final position for 100~ms prior to trap shut off.
\label{f:roughness}}
\end{center}
\end{figure}

Because the perturbations to the magnetic potential above a single
wire are due to its geometric deformations, one expects such
perturbations to increase as the trap center is brought closer to
the surface of the microchip. We observed longitudinal
fragmentation of the BEC as the atoms were brought to within
150~$\mu$m of the surface (Fig.~\ref{f:roughness}). At 55~$\mu$m
from the surface the potential developed axial variations with a
characteristic length scale of $100-150$~$\mu$m. These variations
could be magnetic or gravitational in origin.

In conclusion, we used optical tweezers to load a microfabricated
magnetic trap with $2-3 \times 10^6$ condensate atoms and
subsequently released the atoms into a magnetic waveguide.  Our
setup merges the scientific potential of microfabricated
structures with the proven robustness of conventional BEC
production techniques. In this work, the linear density of atoms
propagating along the waveguide was $10-20$ times larger than the
one-dimensional criterion, which for $^{23}$Na is $n_{1D} \approx
100$~$\mu$m$^{-1}$~\cite{GVL01}. Due to their lack of longitudinal
confinement, waveguides are well suited for creating highly
elongated clouds making them ideal for minimizing mean field
effects~\cite{SFG97} in condensate interferometers and for studies
of one-dimensional physics including phase
fluctuations~\cite{DHR01,PSW00,AAS02} and gases of impenetrable
bosons~\cite{PSW00,DLO01}.

This work was funded by ONR, NSF, ARO, NASA, and the David and
Lucile Packard Foundation.  A.E.L.\ acknowledges additional
support from NSF.  We are indebted to S.~Gupta for contributions
to the apparatus, E.~Tsikata for experimental assistance, the MIT
Microsystems Technology Laboratories for fabricating the microchip
used in this work, M.\ Prentiss and M.\ Vengalattore for valuable
discussions, and A.D.~Cronin and D.~Schneble for critical readings
of the manuscript.


\begin{thebibliography}{23}
\bibitem[{hei()}]{height}
\bibinfo{note}{Analogously, the trajectory will vary in a plane normal to the
  microchip for variations in the wire height.}
\end{thebibliography}
\end{document}